\documentclass{article}

\usepackage[nonatbib,preprint]{neurips_2024}

\usepackage[utf8]{inputenc} 
\usepackage[T1]{fontenc}    
\usepackage{hyperref}       
\usepackage{url}            
\usepackage{booktabs}       

\usepackage{colortbl}

\def\redc{\cellcolor[HTML]{FF999A}}
\def\orangec{\cellcolor[HTML]{FFCC99}}
\def\yellowc{\cellcolor[HTML]{FFF8AD}}
\usepackage{url}
\def\redc{\cellcolor[HTML]{FF999A}}
\def\orangec{\cellcolor[HTML]{FFCC99}}
\def\yellowc{\cellcolor[HTML]{FFF8AD}}
\usepackage{amsfonts}       
\usepackage{nicefrac}       
\usepackage{microtype}      
\usepackage{xcolor}
\usepackage{float}

\usepackage{floatflt}
\usepackage{graphicx}
\usepackage{wrapfig}
\usepackage{colortbl}
\usepackage{pifont}
\usepackage{amsmath}
\usepackage{tikz}
\usepackage{enumitem}

\makeatletter
\newcommand{\printfnsymbol}[1]{%
  \textsuperscript{\@fnsymbol{#1}}%
}
\makeatother

\DeclareMathOperator*{\argmin}{arg\,min}

\usepackage{amsmath}
\usepackage{amssymb}
\usepackage{mathtools}
\usepackage{amsthm}

\theoremstyle{plain}
\newtheorem{theorem}{Theorem}[section]

\newtheorem{corollary}[theorem]{Corollary}
\theoremstyle{definition}

\theoremstyle{remark}

\def\R{\mathbb{R}}

\def\G{\mathcal{G}}

\def\N{\mathcal{N}}

\def\our{NegGS}
\def\ourd{Diff-Gaussian distribution}

\def\redc{\cellcolor[HTML]{FF999A}}
\def\orangec{\cellcolor[HTML]{FFCC99}}
\def\yellowc{\cellcolor[HTML]{FFF8AD}}

\title{\our{}: Negative Gaussian Splatting}

\author{%
  Artur Kasymov\thanks{Equal contribution}  \ 
    \\
    Jagiellonian University\\
  \And
  Bartosz Czekaj\printfnsymbol{1}\\
    Jagiellonian University\\
  \And
   Marcin Mazur\\
  Jagiellonian University\\
  \And
   Jacek Tabor\\
  Jagiellonian University\\
  \And
  Przemysław Spurek\\
 \thanks{ \texttt{przemyslaw.spurek@uj.edu.pl}}
  Jagiellonian University\\ 
IDEAS NCBR
}

\begin{document}

\maketitle
\vspace{-0.8cm}

\begin{abstract}
One of the key advantages of 3D rendering is its ability to simulate intricate scenes accurately. One of the most widely used methods for this purpose is Gaussian Splatting, a novel approach that is known for its rapid training and inference capabilities. In essence, Gaussian Splatting involves incorporating data about the 3D objects of interest into a series of Gaussian distributions, each of which can then be depicted in 3D in a manner analogous to traditional meshes. It is regrettable that the use of Gaussians in Gaussian Splatting is currently somewhat restrictive due to their perceived linear nature. In practice, 3D objects are often composed of complex curves and highly nonlinear structures. This issue can to some extent be alleviated by employing a multitude of Gaussian components to reflect the complex, nonlinear structures accurately. However, this approach results in a considerable increase in time complexity. This paper introduces the concept of negative Gaussians, which are interpreted as items with negative colors. The rationale behind this approach is based on the density distribution created by dividing the probability density functions (PDFs) of two Gaussians, which we refer to as Diff-Gaussian. Such a distribution can be used to approximate structures such as donut and moon-shaped datasets. Experimental findings indicate that the application of these techniques enhances the modeling of high-frequency elements with rapid color transitions. Additionally, it improves the representation of shadows. To the best of our knowledge, this is the first paper to extend the simple elipsoid shapes of Gaussian Splatting to more complex nonlinear structures. 
\end{abstract}


\section{Introduction}

In recent years, the field of neural network-based rendering has witnessed a proliferation of techniques aimed at depicting hitherto inaccessible perspectives of three-dimensional entities and scenarios. Among these techniques are Neural Radiance Fields (NeRFs)~\cite{mildenhall2020nerf}, which have rapidly gained traction within the computer vision and graphics fields, as evidenced by the growing number of publications on the subject (see, e.g., \cite{Gao2022NeRFNR} and references therein). The capacity of NeRFs to generate high-quality renders has been a significant contributing factor to this surge in interest \cite{muller2022instant, li2023neuralangelo, wang2023adaptive}. Nevertheless, despite the enthusiasm and growing research in this area, the extended training and inference duration remains a persistent issue.

In contrast, the recently developed Gaussian Splatting (GS)~\cite{kerbl20233d} method offers rapid training and real-time rendering capabilities. Its distinctive feature is the representation of 3D objects using Gaussian distributions (referred to as Gaussians), obviating the necessity for the use of neural networks in the process. Consequently, Gaussians can be utilized in a manner analogous to the manipulation of 3D point clouds or meshes, thereby enabling operations such as resizing or repositioning within 3D space. 
Nevertheless, the application of Gaussian distributions to produce shape representations necessitates the use of only linear components. On the other hand, 3D objects exhibit intricate curves and significantly nonlinear structures. One partial solution to this problem is the employment of multiple Gaussian components. In precise terms, the approximation of strongly nonlinear structures is accomplished by the use of several Gaussian distributions, which can be considered a piecewise linear approximation.

\begin{figure}[ht!]
\begin{flushleft}
\centering
 Ground truth  \ \  \ \  \ \  \ \  \ \  \our{}   \ \  \ \  Gaussian Splatting   \ \   \ \ Ground truth  \ \ \ \   \ \ \ \ \our{}  \ \ \ \  Gaussian Splatting \\
\end{flushleft}
\centering
\includegraphics[trim=0 0 0 30,clip,width=0.45\textwidth]{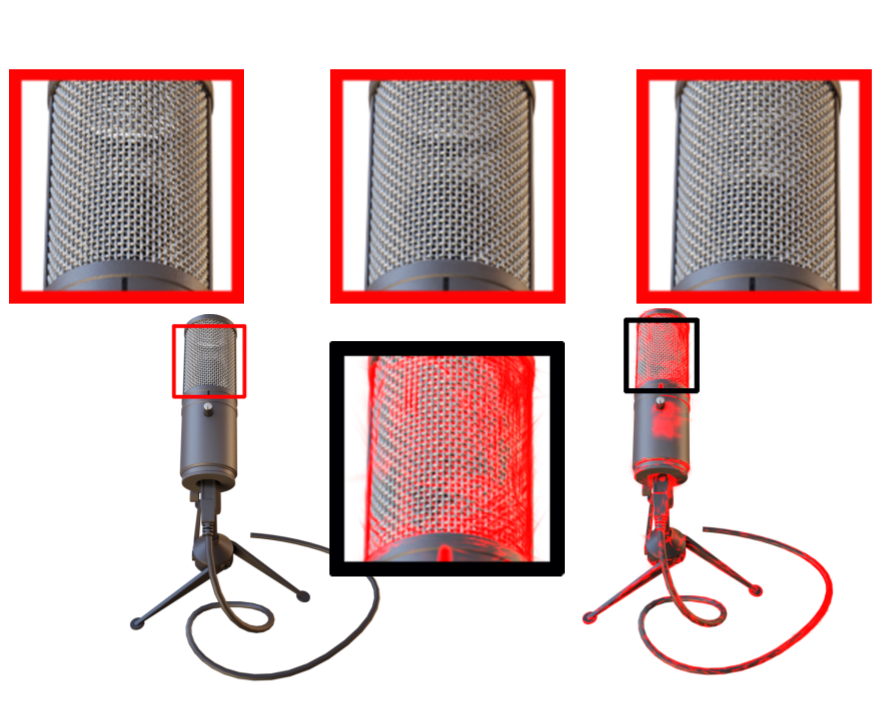} \ \ \ \ \ \ \ \ 
\includegraphics[trim=0 0 0 30,clip,width=0.45\textwidth]{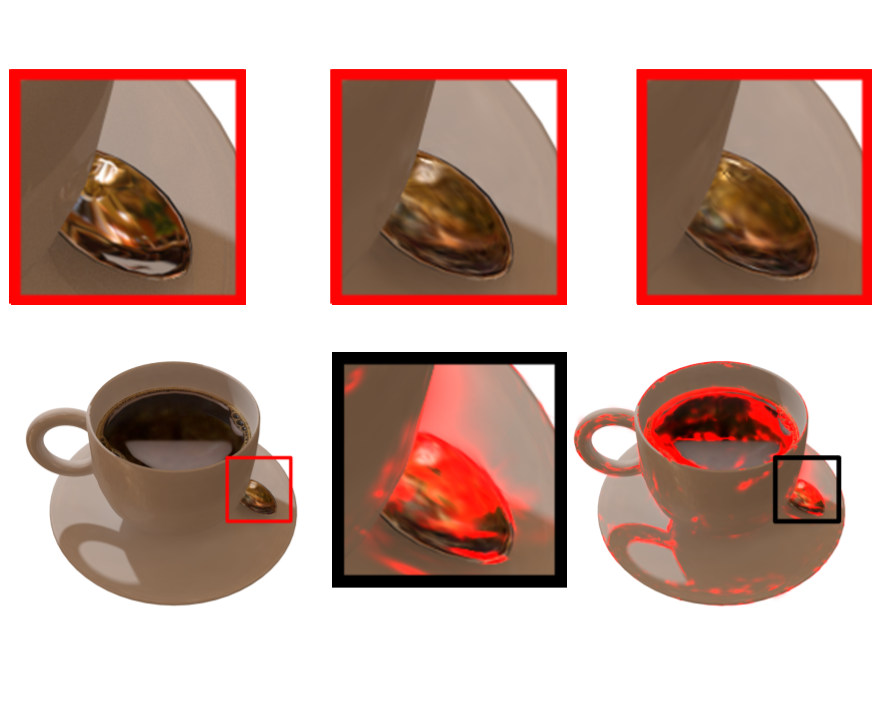}
    \caption{The \our{} (our) model integrates negative Gaussians (Gaussians with negative color) into the GS framework. For visualization purposes, we use a red color to indicate the presence of negative Gaussians, as they cannot be directly rendered. Notably, our method enables the modeling of high-frequency elements with rapidly changing color and light. 
    The presented results were obtained on NeRF Synthetic dataset~\cite{mildenhall2020nerf} and the Shiny Blender dataset~\cite{verbin2022ref}.}
    \label{fig:merf_appro}
\end{figure}



This paper demonstrates the potential advantages of employing more sophisticated probability distributions for the approximation of complicated structures. Our method diverges from the conventional Gaussian framework by incorporating negative components, which we refer to as negative Gaussians, interpreted as items (ellipsoids) with negative colors. This extension allows for the representation of more intricate shapes, such as donut-shaped and moon-shaped structures. Our approach is founded upon a density distribution that has been derived through the subtraction of probability density functions (PDFs) from two Gaussian distributions, referred to as Diff-Gaussian, which enables the encompassing of a broader range of shapes. Consequently, it is possible to integrate such components into the conventional Gaussian Splatting algorithm with only minor modifications. This results in a novel \our{}\footnote{\url{https://github.com/gmum/NegGS}} model that is capable of approximating complex shapes by incorporating a mixture of positive and negative Gaussians.

\begin{figure}[ht!]
\centering
\our{} (our) for shadows \\[-0.25mm]
    \includegraphics[trim=0 0 0 0 clip,width=0.45\textwidth]{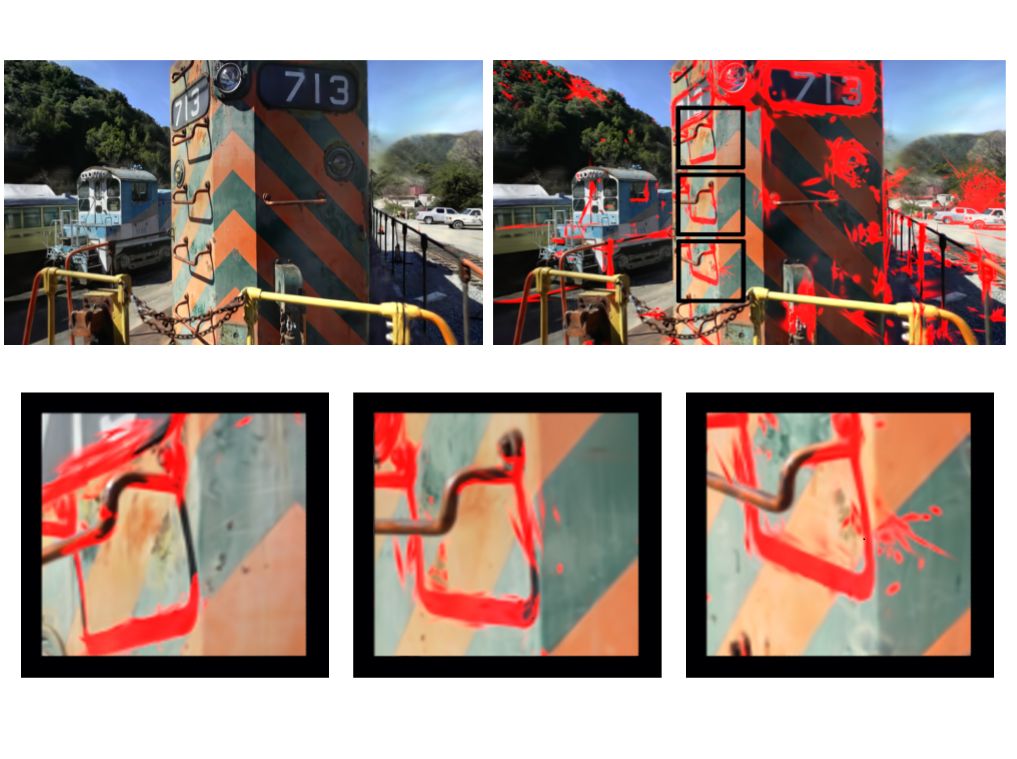} 
    \includegraphics[trim=0 0 0 30,clip,width=0.45\textwidth]{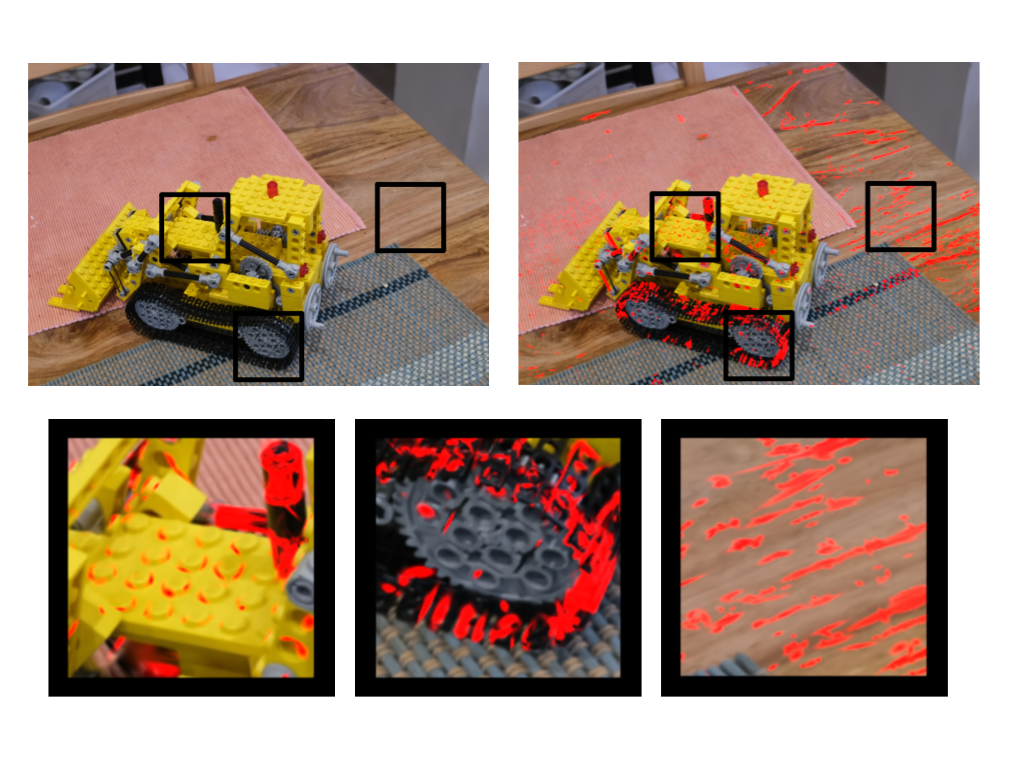}
    \caption{The \our{} (our) model is able to approximate shadows of smaller elements with enhanced precision due to the use of negative Gaussian components (marked in red). The presented results were obtained on the MipNeRF dataset~\cite{barron2022mip}.}
    \label{fig:ex_3}
\end{figure}

The conducted experiments demonstrate that incorporating negative Gaussians into the GS framework facilitates a more accurate approximation of high-frequency regions exhibiting rapid changes in light and shadow (see Figure~\ref{fig:merf_appro}). Furthermore, the shadows of smaller elements can be approximated with enhanced precision (see Figure~\ref{fig:ex_3}). Consequently, the resulting renderings demonstrate enhanced fidelity in specific application contexts when scenes are populated with numerous small elements exhibiting sensitivity to light changes.

Our contribution can be summarized as follows:
\begin{itemize}
    \item we introduce the concept of negative Gaussians (i.e., ellipsoids with negative colors), which improves the approximation of nonlinear structures by Gaussian components, 
    \item we propose a novel \our{} model that incorporates negative Gaussians into the classical Gaussian Splatting algorithm, 
    \item we demonstrate that the proposed method yields superior rendering quality in specific scenarios where scenes comprise numerous small components that exhibit variations in lighting.
\end{itemize}


  

\section{Related work}

The objective of this section is to provide an overview of the Gaussian Splatting algorithm, which serves as the foundation for our proposed solution. Additionally, its main limitations are identified, and selected modifications that are present in the literature are provided.

Gaussian Splatting (GS)~\cite{kerbl20233d} is a method for constructing 3D scenes that employs a set of 3D Gaussians. Each Gaussian is characterized by a collection of parameters, including its mean (position), covariance matrix, opacity, and color, which are encoded using spherical harmonics~\cite{fridovich2022plenoxels,muller2022instant}. Consequently, the GS algorithm operates on the family of tuples
\begin{equation}\label{eq:family_G}
\G_{\mathrm{GS}} = \{ (\N(m_i,\Sigma_i), \alpha_i, c_i) \}_{i=1}^{n},
    \end{equation}
where the means $m_i$, covariance matrices $\Sigma_i$\footnote{In order to stabilize the optimization process, each covariance matrix $\Sigma_i$ is split into a rotation matrix $R_i$ and a scaling matrix $S_i$.}, opacities $\alpha_i$, and colors $c_i$ are to be optimized.
Given the position of a pixel $x$, a sorted list $\G_x\subset\G_{\mathrm{GS}}$ of all tuples with overlapping Gaussians is formed by computing their depths through the viewing transformation $W$. Subsequently, the color $c_x$ of the aforementioned pixel is calculated by alpha compositing, i.e.,
\begin{equation}\label{eq:cx}
c_x = \sum_{\G_x} c_i \alpha'_i \prod_{j=1}^{i-1} (1-\alpha'_j),
    \end{equation}
where the final opacity $\alpha_i'$ is derived from the learned opacity $\alpha_i$ by multiplying it by the 2D Gaussian density $\N(m'_i,\Sigma'_i)$, which is obtained by projecting the respective 3D Gaussian through $W$. 

The GS optimization algorithm employs a cyclical process whereby it renders and then compares the resulting image with the training views derived from the collected data. However, projecting from 3D to 2D can result in misplaced geometry. Consequently, GS optimization requires the capacity to generate, remove, and reposition geometry when it is not correctly placed. The quality of the learned 3D Gaussian covariance matrices is of paramount importance for the achievement of a compact representation, as this allows for the capture of extensive homogeneous regions with a few large anisotropic Gaussians.

The GS procedure is initiated from a few points. The optimization strategy involves the addition of new components and the elimination of redundant ones. At the conclusion of each hundred iterations, GS typically eliminates any Gaussians that exhibit an opacity below a certain threshold. Concurrently, it introduces new Gaussians in empty 3D regions to fill these voids. While the intricacies of the GS optimization process may be challenging to grasp, a robust and efficient implementation and the use of CUDA kernels can facilitate its training in a relatively straightforward manner.

The classical Gaussian splatting technique has undergone numerous modifications in order to facilitate the control of Gaussian components (for further details, see, e.g.,  \cite{wu2024recent} and references therein). As model size is one of the most critical issues to be addressed in GS, the authors of \cite{niedermayr2023compressed,fan2023lightgaussian} propose a number of strategies for GS compression, including the pruning of unnecessary Gaussian components. 
An alternative approach is to use a separate architecture to modulate color, opacity, and shape parameters, as exemplified in \cite{malarzgaussian,yang2024spec}, where the authors employ a neural network that takes positions of Gaussians in order to generate updates contingent on direction of view. This methodology bears a resemblance to the technique utilized in the context of GS scene relighting, as described in \cite{saito2023relightable, gao2023relightable}. For further related results, see the survey paper \cite{chen2024survey}.

\begin{figure}[ht!]
\begin{center} 
    \includegraphics[width=0.2\textwidth, trim=0 0 0 0, clip]{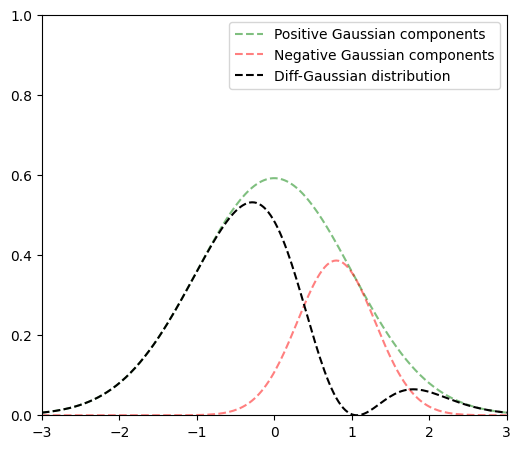}
    \includegraphics[width=0.2\textwidth, trim=0 0 0 0, clip]{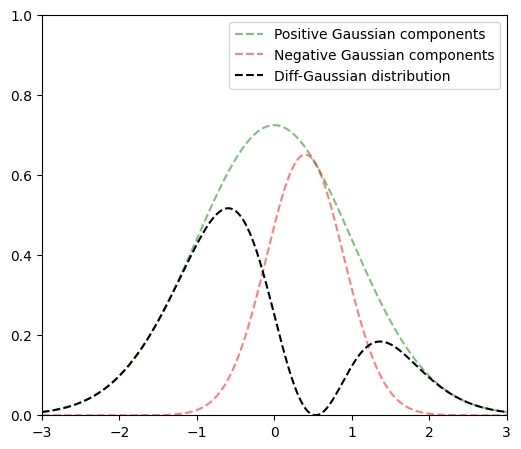}
    \includegraphics[width=0.2\textwidth, trim=0 0 0 0, clip]{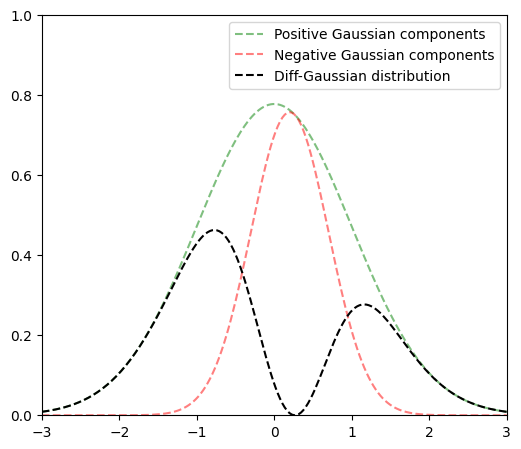}
    \includegraphics[width=0.2\textwidth, trim=0 0 0 0, clip]{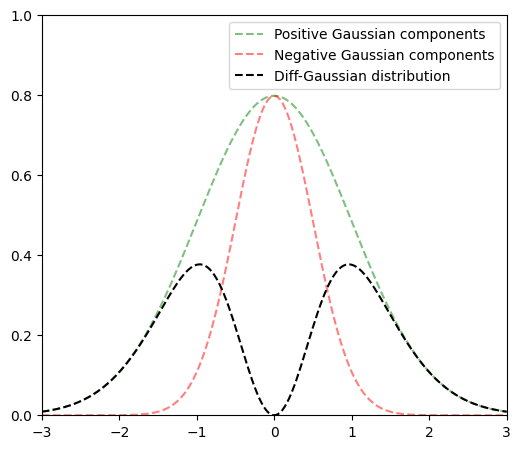}
\end{center} 
\caption{The construction of a 1-dimensional Diff-Gaussian density for different values of parameters. It is of the utmost importance to recognize that the purpose of $c$ is to achieve an equilibrium between the Gaussian components, thus determining the extent to which the negative component can be effectively removed from the positive.}
\label{fig:dist_1d} 
\end{figure}

\begin{figure}[ht!]
\begin{center} 

   \includegraphics[width=0.10\textwidth, trim=30 30 30 30, clip]{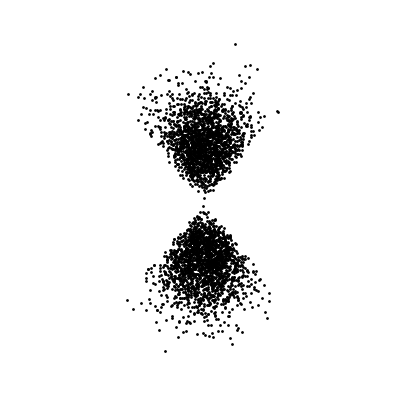}
   \includegraphics[width=0.10\textwidth, trim=30 30 30 30, clip]{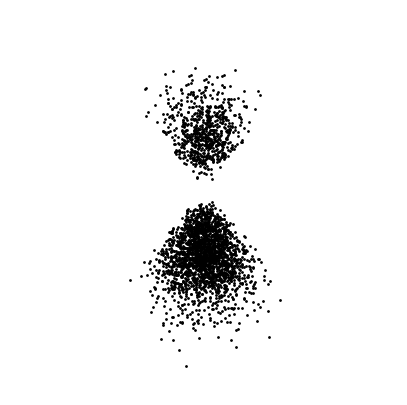}
   \includegraphics[width=0.10\textwidth, trim=30 30 30 30, clip]{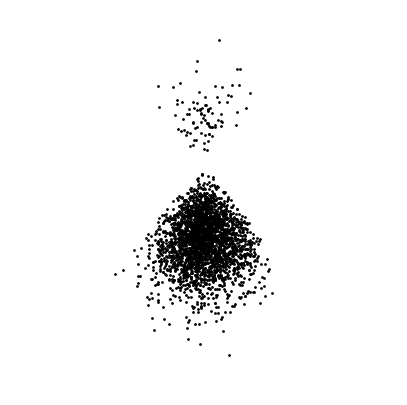}
   \includegraphics[width=0.10\textwidth, trim=30 30 30 30, clip]{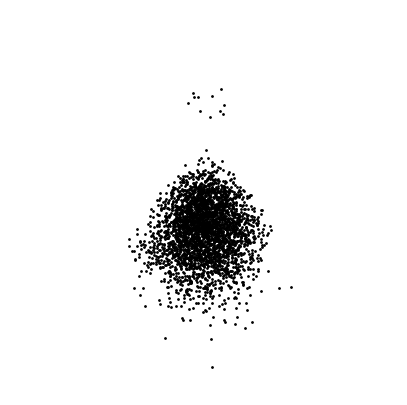}
    \includegraphics[width=0.10\textwidth, trim=30 30 30 30, clip]{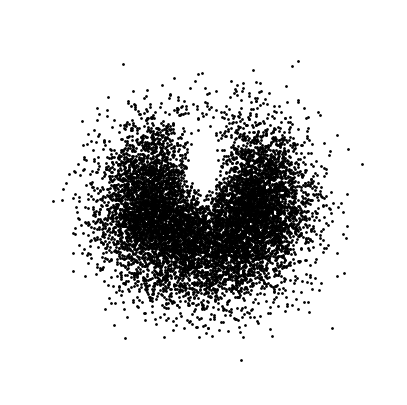}
    \includegraphics[width=0.10\textwidth, trim=30 30 30 30, clip]{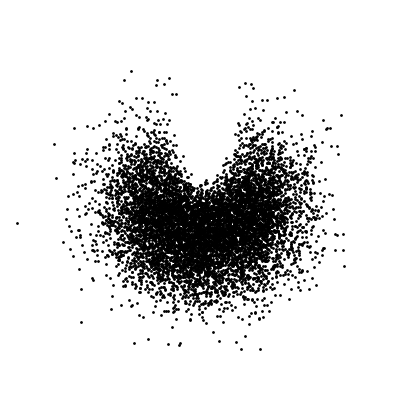}
    \includegraphics[width=0.10\textwidth, trim=30 30 30 30, clip]{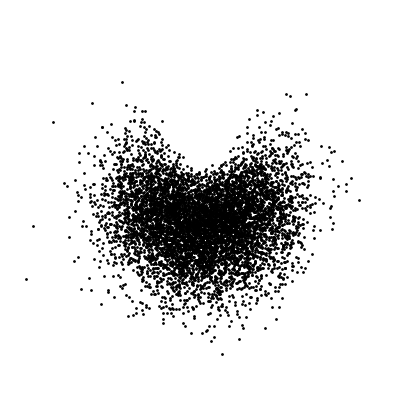}
    \includegraphics[width=0.10\textwidth, trim=30 30 30 30, clip]{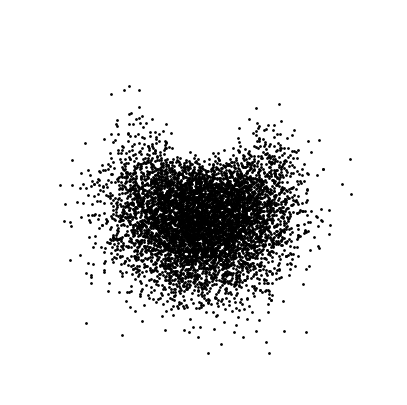}\\
    \includegraphics[width=0.10\textwidth, trim=30 30 30 30, clip]{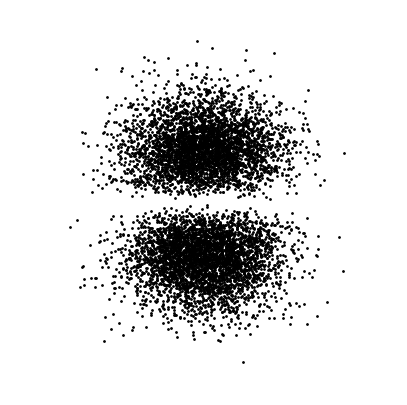}
    \includegraphics[width=0.10\textwidth, trim=30 30 30 30, clip]{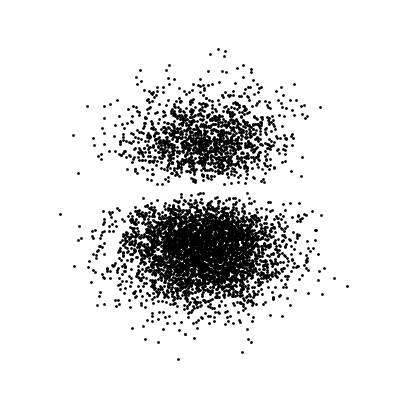}
    \includegraphics[width=0.10\textwidth, trim=30 30 30 30, clip]{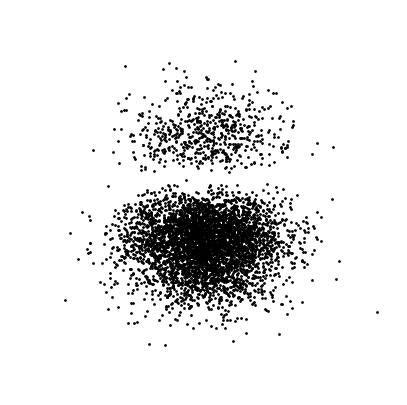}
    \includegraphics[width=0.10\textwidth, trim=30 30 30 30, clip]{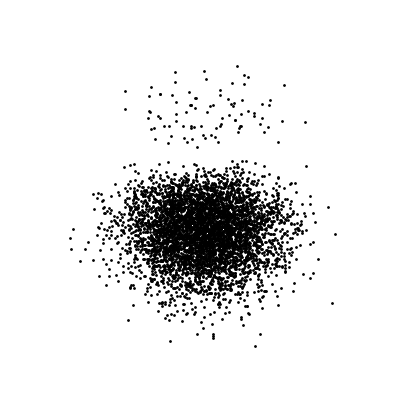}  
    \includegraphics[width=0.10\textwidth, trim=30 30 30 30, clip]{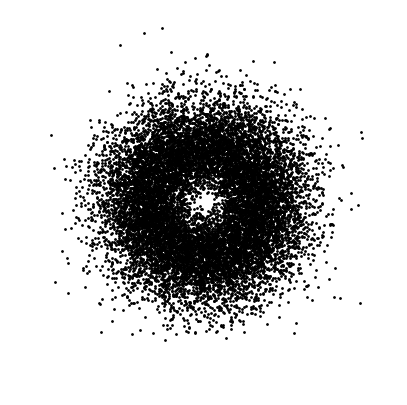}
    \includegraphics[width=0.10\textwidth, trim=30 30 30 30, clip]{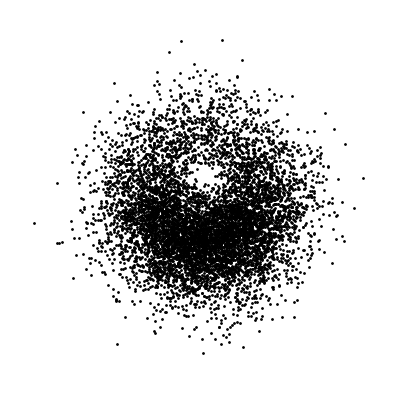}
    \includegraphics[width=0.10\textwidth, trim=30 30 30 30, clip]{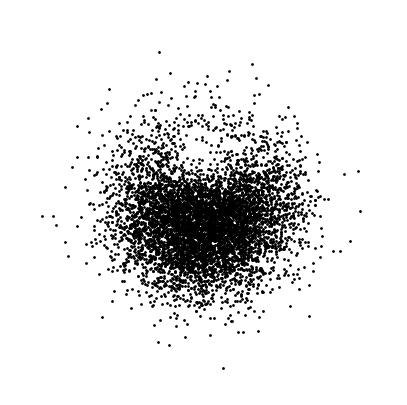}
    \includegraphics[width=0.10\textwidth, trim=30 30 30 30, clip]{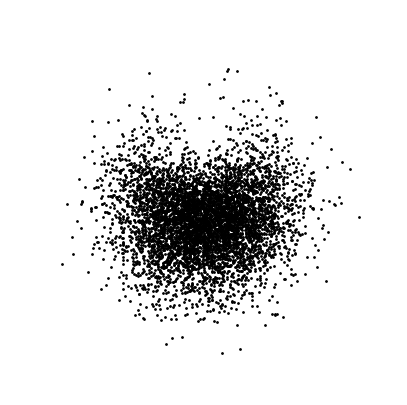}  
    
\end{center} 
\caption{Selected samples derived from various 2-dimensional Diff-Gaussian distributions. Note that a variety of forms are observed, including those resembling an ellipse, a donut, a moon, and a two-part structure.}
\label{fig:dist_2d} 
\end{figure}


To the best of our knowledge, there is no work that extends Gaussian Splatting towards incorporating more sophisticated distributions. Conversely, deviating significantly from the original Gaussian structure would be ineffective due to the complexity of implementation and optimization. In the following section, we propose a novel Gaussian-based family of multivariate distributions that provides a foundation for such an extension of the GS framework.

\section{\ourd{}}
This section introduces the Diff-Gaussian distribution, which provides motivation for integrating negative Gaussians into the original Gaussian Splatting model.

Consider two $n$-dimensional Gaussian distributions, $\N(m_0,\Sigma_0)$ and $\N(m_1,\Sigma_1)$, with respective means vectors $m_i$, covariance matrices $\Sigma_i$ and PDFs $f_{m_i,\Sigma_i}$, where $i\in \{1,2\}$. Define a parameterized family of functions:
\begin{equation}\label{eq:family}
f_{c}(x)=\frac{1}{1-c}\left(f_{m_0,\Sigma_0}(x)-cf_{m_1,\Sigma_1}(x)\right) 
    \end{equation}
for  $0\leq c<1$. The following theorem establishes a condition under which the functions given in Equation~\eqref{eq:family} can be identified as densities, thereby explicitly determining the underlying probability distributions.
\begin{theorem}\label{thm:densities}
Assume that
\begin{equation}\label{eq:assumption}
c\leq \min_{x} \frac{f_{m_0,\Sigma_0}(x)}{f_{m_1,\Sigma_1}(x)}.
\end{equation}
Then the function $f_{c}$ is a density, i.e., it is nonnegative and integrates to 1.
\end{theorem}
The proof of the aforementioned theorem can be derived through straightforward calculations and is therefore omitted here.

In Theorem \ref{thm:densities}, it was demonstrated that Equation~\eqref{eq:family} defines a family of multivariate continuous distributions. We henceforth refer to these as Diff-Gaussians, since each of them is built of two Gaussian components, a positive one (represented by $\N(m_0,\Sigma_0)$) and a negative one (represented by $\N(m_1,\Sigma_1)$). To gain insight into the construction of a Diff-Gaussian density, Figure~\ref{fig:dist_1d} illustrates the respective 1-dimensional examples obtained for different values of the parameters that satisfy the condition given in Equation~\eqref{eq:assumption}. It is important to note that the role of $c$ is to balance the input Gaussian components in order to ascertain the extent to which the negative component can be removed from the positive. Therefore, it may be of interest to explore a situation in which $c$ achieves its optimal value (i.e., when we reach equality in Equation~\eqref{eq:assumption}). This is a case of the following theorem.
\begin{theorem}\label{thm:optimal_c}
Assume that $\Sigma_0-\Sigma_1$ is a positive semi-definite matrix, i.e., $x^T(\Sigma_0-\Sigma_1)x\geq 0$ for every $x$\footnote{This means that the eigenvalues of $\Sigma_0-\Sigma_1$ are all non-negative. (Note that these are real numbers as the matrix is symmetric.)}. Then
\begin{equation}\label{eq:optimal_c}
c_{\text{opt}}:= \min_{x} \frac{f_{m_0,\Sigma_0}(x)}{f_{m_1,\Sigma_1}(x)}=\frac{f_{m_0,\Sigma_0}(w)}{f_{m_1,\Sigma_1}(w)}
\end{equation}
if and only if $w$ is a vector satisfying
\begin{equation}\label{eq:w}
\Sigma_1^{-1}(w-m_1)=\Sigma_0^{-1}(w-m_0).
\end{equation}
\end{theorem}
\proof By Equation~\eqref{eq:optimal_c}, $w$ is an argument for which the function $f_{m_0,\Sigma_0}(x)/f_{m_1,\Sigma_1}(x)$ reaches its minimum. Hence, we can make the following calculations:
\begin{equation}
 \begin{array}{r@{\;}c@{\;}l}
 w & = & \displaystyle \argmin_x \frac{f_{m_0,\Sigma_0}(x)}{f_{m_1,\Sigma_1}(x)} = \argmin_x \left(\log f_{m_0,\Sigma_0}(x) - \log f_{m_1,\Sigma_1}(x)\right)\\[4mm]
 & = & \displaystyle \argmin_x \left((x-m_1)^T\Sigma_1^{-1}(x-m_1)-(x-m_0)^T\Sigma_0^{-1}(x-m_0)\right)\\[4mm]
 & = & \displaystyle \argmin_x \left(x^T(\Sigma_1^{-1}-\Sigma_0^{-1})x-2(m_1^T\Sigma_1^{-1}-m_0^T\Sigma_0^{-1})x\right).
 \end{array}
\end{equation}
From the above, it is evident that we are seeking any $w$ that minimizes the quadratic function $q(x)=x^TAx+2b^Tx$, where $A=\Sigma_1^{-1}-\Sigma_0^{-1}$ and $b=\Sigma_0^{-1}m_0-\Sigma_1^{-1}m_1$. Consequently, since $A$ is a positive semi-definite matrix, it follows that a necessary and sufficient condition for $w$ is that it be a solution of Equation~\eqref{eq:w}. (This can be proven by direct analysis of the gradient and Hessian of $q$ or by reference to, e.g., Theorem~7 in \cite{jankovic2005quadratic}.) 
\endproof

\begin{figure}[ht!]
\centering
\begin{flushleft}
\centering
 \ \  \ \ \ \  Ground truth    \ \  \ \ \ \  \our{}    \ \  Gaussian Splatting   \ \   \ \ Ground truth  \ \ \ \   \ \ \ \ \our{}  \ \ \ \  Gaussian Splatting \\
\end{flushleft}
   \includegraphics[trim=0 0 0 30,clip,width=0.43\textwidth]{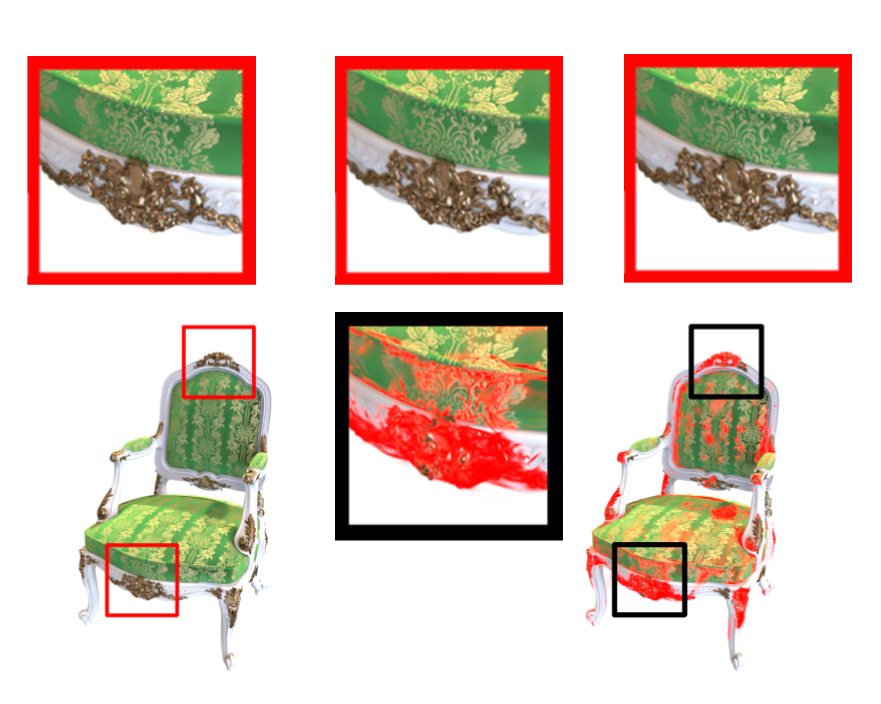} \qquad
    \includegraphics[trim=0 0 0 30,clip,width=0.43\textwidth]{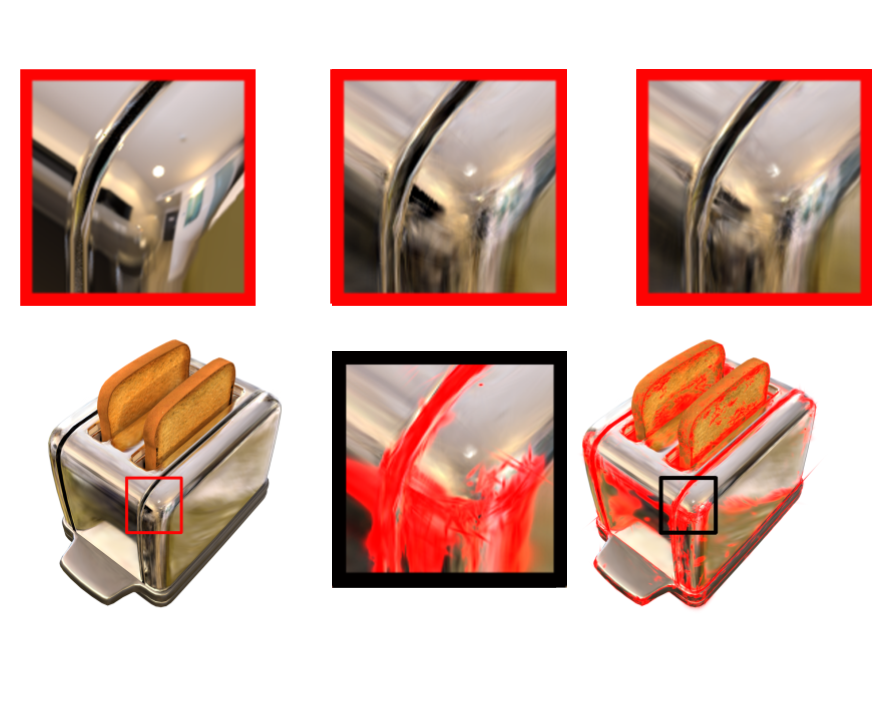} 
    \caption{Qualitative comparison of the \our{} (our) model with Gaussian Splatting. It is noteworthy that our method allows for the modeling of high-frequency elements with rapidly changing color and light due to incorporating negative Gaussian components (marked in red). The presented results were obtained on the NeRF Synthetic dataset~\cite{mildenhall2020nerf} (top) and NeRF Synthetic dataset~\cite{mildenhall2020nerf} (bottom) datasets.}
    \label{fig:ex_2}
\end{figure}

It can be observed from Equation~\eqref{eq:optimal_c} that whenever the set of all vectors $w$ under consideration in Theorem~\ref{thm:optimal_c} is nonempty, it consists of roots of the function $f_{c_{\textit{opt}}}$ (for $c_{\text{opt}}$ provided in Equation~\eqref{eq:optimal_c}), which is a density of the respective Diff-Gaussian distribution (with the exception of the case when $\Sigma_0=\Sigma_1$, in which we have $c_{\textit{opt}}=1$). Moreover, according to Equation~\eqref{eq:w}, this set exhibits the structure of an affine space whose dimension is contingent upon the relation between the covariance matrices. For instance, if $\Sigma_0 - \Sigma_1$ is a positive-definite matrix, i.e., $x^T(\Sigma_0 - \Sigma_1)x>0$ for every $x\neq 0$\footnote{This means that the eigenvalues of $\Sigma_0-\Sigma_1$ are all positive.}, then this set is composed by a single vector (this is due to the fact that in this case the matrix $\Sigma_1^{-1} - \Sigma_0^{-1}$ is also positive-definite and thus invertible). To gain further insight into this dependence, in the following corollary we proceed with a diagonal form of $\Sigma_0$ and $\Sigma_1$, although we emphasize that this does not lead to a loss of generality.

\begin{corollary}\label{cor:diagonal}
Assume that the covariance matrices are diagonal with respect to some linear basis of $n$-dimensional space, i.e., \begin{equation}\label{eq:diagcov}
\Sigma_0=\operatorname{diag}(\sigma_1^2,\ldots,\sigma_n^2) \; \text{ and }\;  \Sigma_1=\operatorname{diag}(h_1\sigma_1^2,\ldots,h_n\sigma_n^2),
\end{equation}
where $\sigma_i>0$, $0<h_i\leq 1$ for all $i\in \{1,\ldots, n\}$ (to ensure that $\Sigma_0-\Sigma_1$ is positive semi-definite), and $h_j<1$ for some $j\in \{1,\ldots, n\}$ (to ensure that $c_{\textit{opt}}<1$). Then a vector $w$ is a root of the Diff-Gaussian density $f_{c_{\text{opt}}}$ if and only if for every $i\in \{1,\ldots, n\}$ one of the following conditions hold:
\begin{itemize}
    \item[(i)] either $w_i=(m_{1,i}-h_im_{0,i})/(1-h_i)$ if $h_i<1$,
    \item[(ii)] or $w_i\in \R$ if $h_i=1$ and $m_{1,i}=m_{0,i}$,
\end{itemize}
where $w_i$, $m_{0,i}$, and $m_{1,i}$ denote the respective coordinates of $w$, $m_0$, and $m_1$.
\end{corollary}
The proof of the aforementioned corollary can be derived directly from Equation~\eqref{eq:w}.

\begin{table}[ht!]
\centering
    \caption{Quantitative comparison of \our{} (our) with the state-of-the-art methods, conducted on the NeRF Synthetic dataset~\cite{mildenhall2020nerf}. The PSNR, SSIM, and LPIPS evaluation metrics were employed, and the experimental benchmark proposed in \cite{kerbl20233d} was utilized. The cells with the three highest scores are colored red, orange, and yellow, respectively.}
    \label{tab:nerf}
    {\small 
    \begin{tabular}{c@{\;}c@{\;}c@{\;}c@{\;}c@{\;}c@{\;}c@{\;}c@{\;}c@{\;}c}
   \multicolumn{10}{c}{PSNR $\uparrow$} \\
\toprule      
& Chair & Drums & Lego & Mic & Materials & Ship & Hotdog & Ficus & Avg. \\ \midrule

NeRF & 33.00 & 25.01 & 32.54 & 32.91 & \yellowc 29.62 & 28.65 & 36.18 & \orangec 30.13 & 31.01 \\ 
VolSDF & 30.57 & 20.43 & 29.46 & 30.53 & 29.13 & 25.51 & 35.11 & 22.91 & 27.96 \\ 
Ref-NeRF & \yellowc 33.98 & \yellowc 25.43 & \yellowc 35.10 & \yellowc 33.65 & 27.10 & \yellowc 29.24 & \yellowc 37.04 & \yellowc 28.74 & \yellowc 31.29 \\ 
ENVIDR & 31.22 & 22.99 & 29.55 & 32.17 & 29.52 & 21.57 & 31.44 & 26.60 & 28.13 \\ 
\midrule
GS & \orangec 35.82 & \redc 26.17 & \redc 35.69 & \orangec 35.34 & \orangec 30.00 & \redc 30.87 & \redc 37.67 & \redc 34.83 & \orangec 33.30 \\ 
\our{} (our) & \redc 35.88 & \orangec 26.14 & \orangec 35.68 & \redc 36.15 & \redc 30.12 & \orangec 30.50 & \orangec 37.55 & \redc 34.83 & \redc 33.36

\\ \bottomrule
\\ 
  \multicolumn{10}{c}{SSIM $\uparrow$} \\
\toprule
& Chair & Drums & Lego & Mic & Materials & Ship & Hotdog & Ficus & Avg. \\ \midrule
NeRF & 0.967 & 0.925 & 0.961 & 0.980 & 0.949 & 0.856 & \yellowc 0.974 & \orangec 0.964 & 0.947 \\
VolSDF  & 0.949 & 0.893 & 0.951 & 0.969 &  0.954 & 0.842 & 0.972 & 0.929 & 0.932 \\
Ref-NeRF & \yellowc 0.974 & \yellowc 0.929 & \yellowc 0.975 & 0.983 & 0.921 & \yellowc 0.864 & \orangec 0.979 & \yellowc 0.954 & 0.947 \\
ENVIDR & \orangec 0.976 & \orangec 0.930 & 0.961 & \yellowc 0.984 & \redc 0.968 & 0.855 & 0.963 & \redc 0.987 & \yellowc 0.956 \\
\midrule
GS & \redc 0.987 & \redc 0.954 & \redc 0.983 & \orangec 0.991 & \yellowc 0.960 & \redc 0.907 & \redc 0.985 & \redc 0.987 &  \orangec 0.969 \\
\our{} (our)  &  \redc 0.987 & \redc 0.954 & \orangec 0.982 & \redc 0.992 & \orangec 0.962 & \orangec 0.906 &\redc 0.985 & \redc0.987 & \redc 0.970
\\ \bottomrule
\\
    \multicolumn{10}{c}{LPIPS $\downarrow$ }\\
\toprule
& Chair & Drums & Lego & Mic & Materials & Ship & Hotdog & Ficus & Avg. \\ \midrule
NeRF & 0.046 & 0.091 & \yellowc 0.050 &  0.028 & 0.063 & 0.206 & 0.121 & \orangec 0.044 & 0.081 \\
VolSDF & \yellowc 0.056 & 0.119 & 0.054 & 0.191 & \yellowc 0.048 & 0.191 & 0.043 & 0.068 & 0.096 \\
Ref-NeRF & \orangec 0.029 & 0.073 & \yellowc \orangec 0.025 & \orangec 0.018 & 0.078 & \yellowc 0.158 & \yellowc 0.028 & 0.056 & \orangec 0.058 \\
ENVIDR & \yellowc 0.031 & \yellowc 0.080 & 0.054 &  \yellowc 0.021 & \orangec 0.045 & 0.228 & 0.072 & \redc 0.010 & \yellowc 0.067 \\ 
\midrule
GS & \redc 0.012 &  \redc 0.037 & \redc 0.016 & \redc 0.006 & \redc 0.034 & \orangec 0.106 & \redc 0.020 & \orangec 0.012 & \redc 0.030 \\
\our{} (our) & \redc 0.012 &  \orangec 0.038 & \redc 0.016 &\redc 0.006 &\redc 0.034 & \redc0.105 & \orangec 0.021 & \orangec 0.012 & \redc 0.030 
\\ \bottomrule
\end{tabular}
    }
\end{table}

Corollary~\ref{cor:diagonal} demonstrates that it is feasible to impose a null density of a Diff-Gaussian distribution on a $k$-dimensional affine space, where $k\in \{0,\ldots,n-1\}$, by aligning the means and variances of its Gaussian components along respective 1-dimensional directions. This implies that the use of such distributions to approximate complex real-world scenes may be more effective than the use of standard Gaussians (as exemplified by the classical Gaussian Splatting algorithm), given that a single Diff-Gaussian density can encompass a broader range of shapes. Illustrative 2D samples are presented in Figure~\ref{fig:dist_2d}. It is crucial to recognize that, in addition to Gaussian elipses, various other shapes can be attained, including those that resemble donuts or moons, as well as more incoherent forms.

The following section presents a method for modifying the Gaussian Splatting algorithm to incorporate the concept of a Diff-Gaussian distribution.

\section{Negative Gaussian Splatting}

In this section, we present the concept of our Negative Gaussian Splatting (\our{}) model, which is motivated by the observed superiority of Diff-Gaussians over standard Gaussians in covering complex nonlinear structures.

\begin{table}[]
  \caption{Quantitative comparison of \our{} (our) with the state-of-the-art methods, conducted on the Mip-NeRF360 \cite{barron2022mip}, Tanks and Temples \cite{knapitsch2017tanks}, and Deep Blending \cite{hedman2018deep} datasets. The PSNR, SSIM, and LPIPS evaluation metrics were employed, and the experimental benchmark proposed in \cite{kerbl20233d} was utilized. The cells with the three highest scores are colored red, orange, and yellow, respectively.}
    \label{tab:real_data}
{
\begin{center}
    \begin{tabular}{@{}c ccc ccc ccc @{}}
    &  \multicolumn{3}{c}{Mip-NeRF360} & \multicolumn{3}{c}{Tanks and Temples} & \multicolumn{3}{c}{Deep Blending} \\
    \toprule
         & SSIM $\uparrow$& PSNR $\uparrow$& LPIPS $\downarrow$& SSIM $\uparrow$& PSNR $\uparrow$& LPIPS $\downarrow$ & SSIM $\uparrow$& PSNR $\uparrow$& LPIPS $\downarrow$ \\ \midrule

Plenoxels & 0.626 & 23.08 &  0.719 & 0.379& 
21.08& 0.795 & 
0.510& 23.06& 0.510
\\
INGP-Base & 0.671 & 25.30 & 0.371 & 0.723 & 21.72 & 0.330 & 0.797 & 23.62 & 0.423
\\
INGP-Big & 0.699 & 25.59 & 0.331 & 0.745 & 21.92 & 0.305 & 0.817 & 24.96 & 0.390
\\
M-NeRF360 &\yellowc  0.792 & \redc 27.69 & \yellowc 0.237 & 0.759 & \yellowc 22.22 & \yellowc 0.257 & \orangec  0.901 &\yellowc 29.40 & \orangec   0.245
 \\
 \midrule
GS-7K & 0.770 & 25.60 & 0.279 & 0.767 & 21.20 & 0.280 & 0.875 & 27.78 & 0.317
\\
GS-30K & \redc 0.815 & \yellowc 27.21 & \redc 0.214  & \orangec 0.841 &  \orangec 23.14 &  \orangec 0.183 & \redc 0.903 &  \orangec  29.41 & \redc 0.243\\

\our{}-7K (Our) & 0.765 & 25.87 & 0.287 & \yellowc  0.775 & 21.66 & 0.270 & 0.875 & 28.11 & 0.315   \\
\our{}-30K (Our) &  \orangec 0.812 &  \orangec 27.39 &  \orangec 0.219  & \redc 0.844 & \redc 23.61 & \redc 0.179 & \yellowc 0.900 & \redc 29.55 & \yellowc 0.247  

 \\ \bottomrule
     
    \end{tabular}
    \end{center}
    }
    
\end{table}

In general, the primary distinction between the \our{} and the classical Gaussian Splatting~\cite{kerbl20233d} is the incorporation of negative Gaussians (i.e., ellipsoids with negative colors) as components of a family of tuples that are optimized to represent a given 3D scene. To be precise, in Equation~\eqref{eq:family_G}, any $c_i$ may be a negative color value (i.e., such that $-c_i$ represents the respective color). Accordingly, the optimization algorithm, which is a derivative of the GS method, operates on a joint family
\begin{equation}\label{eq:family_GN}
\G_{\mathrm{NeGS}} = \G_{\mathrm{GS}}\cup \G_{\mathrm{-GS}},
\end{equation}
wherein $\G_{\mathrm{GS}}$ and $\G_{\mathrm{-GS}}$ comprise tuples characterized by positive and negative values of $c_i$, respectively.

It should be noted that the aforementioned extension can be interpreted as the use of Diff-Gaussians, rather than standard Gaussians, in the GS framework. This affords a greater degree of flexibility in the generation of desired shape representation and color modulation, as two Gaussians sharing identical parameters and inverse color reduce each other. It should also be noted that additional strategies are employed to control the number of Gaussians. Specifically, Gaussians with low opacity are reduced. Conversely, new Gaussians are introduced into empty 3D regions, thereby enabling the number of Gaussians to be dynamically adapted. However, our experiments demonstrate that the initial number of negative Gaussians is of critical importance for the efficacy of the optimization procedure. Consequently, the numbers of positive and negative components are treated as hyperparameters, which are adapted through a grid search procedure during the cross-validation stage. It is observed that, in the majority of cases, the number of negative Gaussians required is significantly less than that of positive Gaussians. This is indicative of the fact that negative Gaussians are employed to enhance the visibility of high-frequency areas, with the elements comprising these areas being particularly sensitive to changes in light conditions. 

In essence, the \our{} optimization procedure allows for the application of simple corrections in the colors of small pieces of represented objects (by involving appropriate components with negative colors), eliminating the need for the use of more (positive) Gaussians instead. To illustrate this point, one may consult Figure~\ref{fig:ex_3}, which suggests that minor shades can be efficiently covered by the use of negative Gaussians (marked in red), thereby demonstrating the superiority of our solution in comparison to Gaussian Splatting.

The following section presents the experimental results obtained using our proposed algorithm on a selection of synthetic and real-world datasets, in comparison to those produced by the leading state-of-the-art methods, including the original GS procedure.

\begin{table}[ht!]
\centering
        \caption{Quantitative comparison of \our{} (our) with the state-of-the-art methods, conducted on the Shiny Blender dataset \cite{verbin2022ref}. The PSN, SSIM, and LPIPS evaluation metrics were employed, and the experimental benchmark proposed in \cite{verbin2022ref} was utilized. The cells with the three highest scores are colored red, orange, and yellow, respectively.}
    \label{tab:shiny}
\begin{tabular}{c@{\;}c@{\;}c@{\;}c@{\;}c@{\;}c@{\;}c@{\;}c}
    \multicolumn{8}{c}{PSNR $\uparrow$} \\
    \toprule
         & Car & Ball & Helmet & Teapot & Toaster & Coffee & Avg.
 \\ \midrule

NVDiffRec & \orangec 27.98 & 21.77 & 26.97 & 40.44 & 24.31 & 30.74 & 28.70
 \\
NVDiffMC & 25.93 & \orangec 30.85 & 26.27 & 38.44 & 22.18 & 29.60 & 28.88
 \\
Ref-NeRF & \redc  30.41 & 29.14 & \orangec 29.92 & \yellowc 45.19 & \yellowc 25.29 & \redc  33.99 & \orangec 32.32
 \\
NeRO & 25.53 & \yellowc 30.26 & \yellowc 29.20 & 38.70 & \redc 26.46 & 28.89 & 29.84
 \\
ENVIDR & \orangec 28.46 & \redc  38.89 & \redc  32.73 & 41.59 & \orangec 26.11 & 29.48 & \redc  32.88
 \\
 \midrule
GS & 27.24 & 27.69 & 28.32 & \orangec 45.68 & 20.99 & \yellowc 32.32 & \yellowc 30.37  \\
\our{} (our) & 27.27 & 27.74 & 28.18 & \redc 45.70 & 21.00 & \orangec 32.35 & \yellowc 30.37 \\
\bottomrule\\
    \multicolumn{8}{c}{SSIM $\uparrow$} \\
    \toprule
           & Car & Ball & Helmet & Teapot & Toaster & Coffee & Avg.
 \\ \midrule
NVDiffRec & \redc 0.963 & 0.858 & 0.951 & \orangec 0.996 & \yellowc 0.928 & \redc 0.973 & 0.945\\
NVDiffMC & 0.940 & 0.940 & 0.940 & \yellowc 0.995 & 0.886 & 0.965 & 0.944\\
Ref-NeRF & \yellowc 0.949 & \yellowc 0.956 & \yellowc 0.955 & \yellowc 0.995 & 0.910 &  \orangec 0.972 & \yellowc 0.956\\
NeRO & \yellowc 0.949 & \orangec 0.974 & \orangec 0.971 & \yellowc 0.995 & \orangec 0.929 & 0.956 & \orangec 0.962\\
ENVIDR & \orangec 0.961 & \redc 0.991 & \redc  0.980 & \orangec 0.996 & \redc 0.939 & 0.949 & \redc 0.969\\
\midrule
GS  & 0.930 & 0.937 & 0.951 & \orangec 0.996 & 0.895 & \yellowc 0.971 & 0.947\\
\our{} (our) & 0.931 & 0.938 & 0.951 & \redc 0.997 & 0.895 & \orangec 0.972 & 0.947 \\

\bottomrule\\
    \multicolumn{8}{c}{LPIPS $\downarrow$ }\\
    \toprule
           & Car & Ball & Helmet & Teapot & Toaster & Coffee & Avg.
 \\ \midrule
NVDiffRec & \redc 0.045 & 0.297 & 0.118 & \orangec 0.011 & 0.169 & \redc 0.076 & 0.119 \\
NVDiffMC & 0.077 & 0.312 & 0.157 & 0.014 & 0.225 & 0.097 & 0.147  \\
Ref-NeRF & 0.051 & 0.307 & 0.087 & 0.013 & \yellowc 0.118 & \yellowc 0.082 &  \yellowc 0.109  \\
NeRO & 0.074 & \orangec 0.094 & \redc 0.050 & \yellowc 0.012 & \redc 0.089 & 0.110 & \redc 0.072  \\
ENVIDR & 0.049 & \redc 0.067 & \orangec 0.051 & \orangec 0.011 & \orangec 0.116 & 0.139 & \redc 0.072  \\
\midrule
GS & \orangec 0.047 & 0.161 & \yellowc 0.079 & \redc 0.007 & 0.126 & \orangec 0.078 & \orangec 0.083  \\
\our{} (our) & \yellowc 0.048 & \yellowc 0.160 & 0.080 & \redc 0.007 & 0.128 & \orangec 0.078 & \orangec 0.083
 \\ \bottomrule
    \end{tabular}
\end{table} 

\section{Experiments}
\label{submission}

\begin{figure}[ht!]
\begin{center} 
    \includegraphics[width=0.31\columnwidth, trim=0 0 0 0, clip]{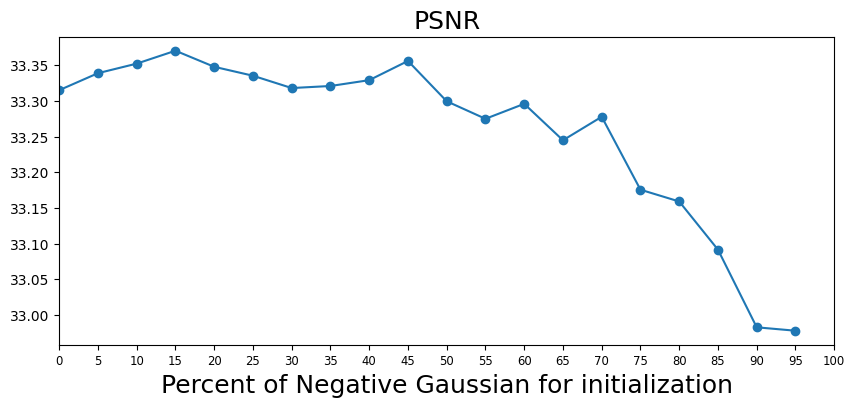}
    \includegraphics[width=0.31\columnwidth, trim=0 0 0 0, clip]{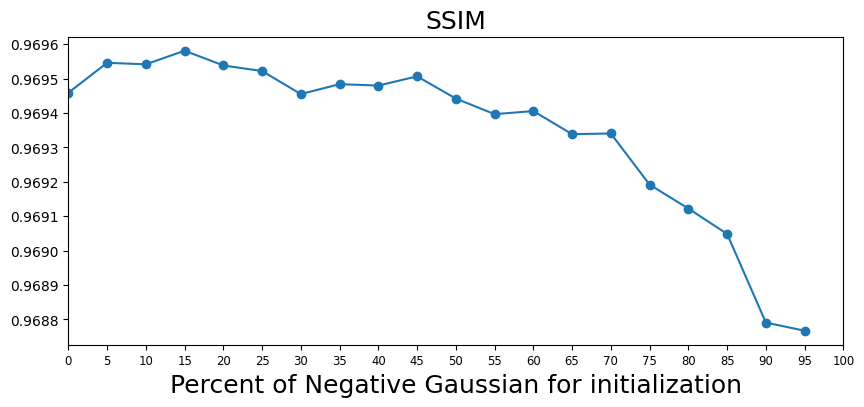}
    \includegraphics[width=0.31\columnwidth, trim=0 0 0 0, clip]{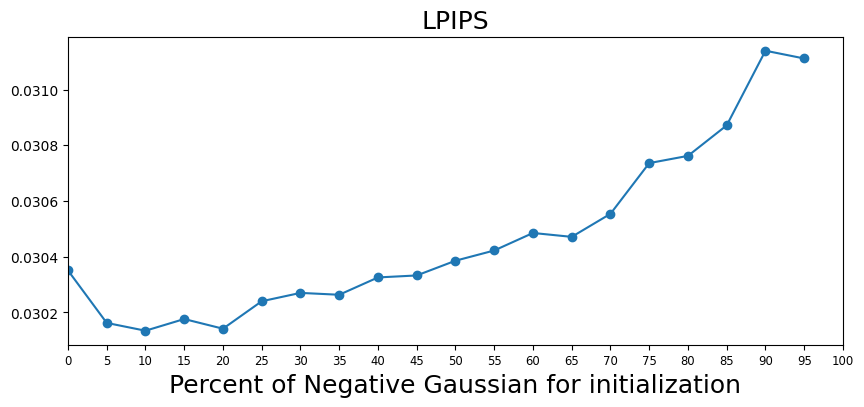}
\end{center} 
\caption{Averaged PSNR$\uparrow$, SSIM$\uparrow$, and LPIPS$\downarrow$ metrics values vs. the proportion of negative Gaussians used during initialization, obtained by \our{} (our) on the NeRF Synthetic dataset~\cite{mildenhall2020nerf}. It can be seen that our method is most effective when about 20\% of the Gaussians used are negative.}
\label{fig:dist_1} 
\end{figure}

\begin{figure}[ht!]
\begin{center} 
    \includegraphics[width=0.75\columnwidth, trim=0 0 0 0, clip]{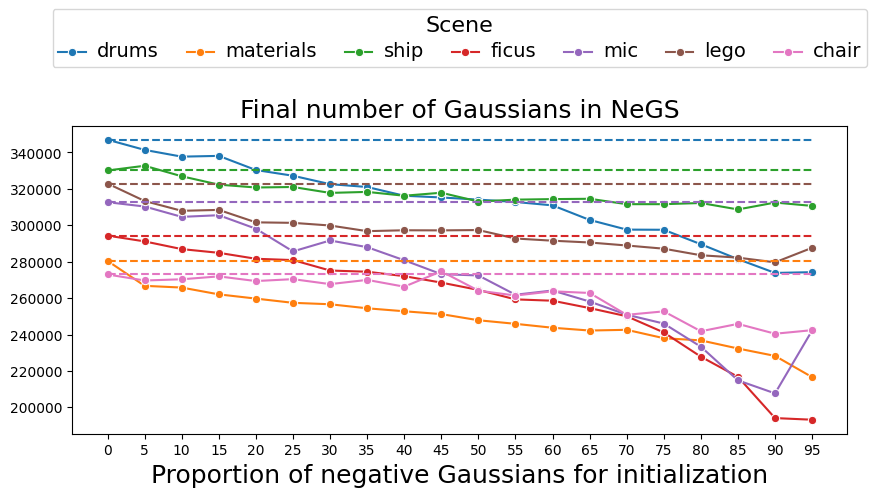}
\end{center}
\caption{The final number of positive and negative Gaussians received by the \our{} training procedure, initialized with different numbers of negative Gaussians. It should be noted that employing negative components allows our method to require fewer total numbers of Gaussians. The presented results were obtained on the NeRF Synthetic dataset~\cite{mildenhall2020nerf}.}
\label{fig:num_gau} 
\end{figure}

In this section, we present the results of comprehensive experiments conducted to assess the performance of the \our{} model across a variety of classical datasets, categorized into three groups:
\begin{itemize}
\item {\em synthetic datasets}, which provide controlled environments for testing,
\item {\em real-world datasets}, which feature diverse and complex scenes,
\item {\em reflection-oriented datasets}, which focus on objects with high reflectivity and transparency.
\end{itemize}
 Our study is intended to provide supplementary evidence in support of previous findings and to present additional quantitative and qualitative results. Specifically, we demonstrate that our model is effective in certain scenarios, particularly in the representation of high-frequency elements with rapid color changes (as illustrated in Figures~\ref{fig:ex_2}), and in the capture of small shadows (as shown in Figure~\ref{fig:ex_3}). Furthermore, we demonstrate that our proposed solution yields comparable or superior quantitative results compared to existing methods (as evidenced in Tables~\ref{tab:nerf}, \ref{tab:real_data}, and \ref{tab:shiny}). Additionally, we provide an ablation study, which verifies the influence of the initial number of negative Gaussian components on the optimization process (see Figures~\ref{fig:dist_1} and \ref{fig:num_gau}).

All experiments were conducted with a consistent set of hyperparameters, identical to those used in GS~\cite{kerbl20233d}, and could be executed on a single NVIDIA RTX 4090 GPU. Notably, our method fully leverages the GS framework implementation, extending it with negative Gaussians tracked via a binary mask, thereby preserving the original GS rendering speed.



%








\paragraph{Experiments on synthetic data} 
Initial tests were conducted using the NeRF Synthetic dataset~\cite{mildenhall2020nerf}, which comprises a variety of objects with intricate geometries and realistic non-Lambertian surfaces. The numerical outcomes are presented in Table~\ref{tab:nerf}, and visual comparisons are shown in Figure~\ref{fig:ex_2}. The results demonstrate that our approach outperformed existing techniques, including Gaussian Splatting.

\paragraph{Experiments on real-world data} Furthermore, \our{} was evaluated on real scenes from the Mip-NeRF360~\cite{barron2022mip} and Deep Blending~\cite{hedman2018deep} datasets, where it demonstrated performance comparable to the GS model (as shown in Table~\ref{tab:real_data}). Nevertheless, on the larger-scale Tanks and Temples dataset, our model exhibited superior capabilities in modeling light reflections and shadows compared to GS and NeRF, as evidenced by the highest scores across most evaluated datasets and metrics (see Table~\ref{tab:real_data}).

\paragraph{Experiments on reflection-oriented data} Finally, experiments were conducted on the Shiny Blender dataset~\cite{verbin2022ref}. Quantitative assessments in Table~\ref{tab:shiny} reveal that the results obtained by \our{} are on par with those from NeRF and GS. Furthermore, our model demonstrates the effective simulation of shadows, light reflections, and transparency in 3D objects, as evidenced in Figure~\ref{fig:merf_appro}.

\paragraph{Ablation study}

The objective of the ablation study was to ascertain the impact of augmenting the initial number of negative Gaussians in \our{}. The outcomes are presented in Figure~\ref{fig:dist_1}, which depicts the PSNR, SSIM, and LPIPS metrics values as a function of the proportion of negative Gaussians employed in the model. It is evident that our method is most effective when approximately 20\% of the Gaussians used are negative. Furthermore, it can be observed that the increasing utilization of negative Gaussians results in a reduction in the total number of Gaussian components, as illustrated in Figure~\ref{fig:num_gau}.

\section{Conclusions}

This work introduces Negative Gaussian Splatting (\our{}), a novel approach in 3D scene rendering. The core concept involves a newly defined Diff-Gaussian distribution, which is capable of modeling nonlinear shapes through the joint use of positive and negative Gaussian components. Drawing inspiration from the Diff-Gaussian distribution, we propose to incorporate negative Gaussian components that utilize negative colors. As demonstrated in the paper, our method is highly efficient in training and introduces only minor modifications to the GS framework. Moreover, \our{} can capture high-frequency details with rapid color changes and improve shadow representation. To the best of our knowledge, our work is the first to extend Gaussian Splatting toward adapting basic ellipsoid forms into more intricate nonlinear configurations.

\paragraph{Limitation}

Our \our{} model demonstrates superior performance when applied to specific areas of 3D scenes that contain high-frequency details with rapid color changes. Otherwise, when used on shapes with simple geometry, the results are comparable to those achieved by Gaussian Splatting. Furthermore, the Diff-Gaussian distribution is not explicitly utilized in our approach; rather, negative Gaussians are incorporated into the GS framework, which yields a similar effect. It is not immediately evident how to implement the direct use of the Diff-Gaussian distributions in our proposed solution.

\section*{}

\bibliographystyle{unsrt}

\begin{thebibliography}{10}

\bibitem{mildenhall2020nerf}
Ben Mildenhall, Pratul~P. Srinivasan, Matthew Tancik, Jonathan~T. Barron, Ravi
  Ramamoorthi, and Ren Ng.
\newblock {NeRF: Representing Scenes as Neural Radiance Fields for View
  Synthesis}.
\newblock In {\em ECCV}, 2020.

\bibitem{verbin2022ref}
Dor Verbin, Peter Hedman, Ben Mildenhall, Todd Zickler, Jonathan~T Barron, and
  Pratul~P Srinivasan.
\newblock Ref-nerf: Structured view-dependent appearance for neural radiance
  fields.
\newblock In {\em 2022 IEEE/CVF Conference on Computer Vision and Pattern
  Recognition (CVPR)}, pages 5481--5490. IEEE, 2022.

\bibitem{Gao2022NeRFNR}
Kyle Gao, Yin Gao, Hongjie He, Denning Lu, Linlin Xu, and Jonathan Li.
\newblock Nerf: Neural radiance field in 3d vision, a comprehensive review.
\newblock {\em ArXiv}, abs/2210.00379, 2022.

\bibitem{muller2022instant}
Thomas M{\"u}ller, Alex Evans, Christoph Schied, and Alexander Keller.
\newblock Instant neural graphics primitives with a multiresolution hash
  encoding.
\newblock {\em ACM Transactions on Graphics (ToG)}, 41(4):1--15, 2022.

\bibitem{li2023neuralangelo}
Zhaoshuo Li, Thomas M{\"u}ller, Alex Evans, Russell~H Taylor, Mathias Unberath,
  Ming-Yu Liu, and Chen-Hsuan Lin.
\newblock Neuralangelo: High-fidelity neural surface reconstruction.
\newblock In {\em Proceedings of the IEEE/CVF Conference on Computer Vision and
  Pattern Recognition}, pages 8456--8465, 2023.

\bibitem{wang2023adaptive}
Zian Wang, Tianchang Shen, Merlin Nimier-David, Nicholas Sharp, Jun Gao,
  Alexander Keller, Sanja Fidler, Thomas M{\"u}ller, and Zan Gojcic.
\newblock Adaptive shells for efficient neural radiance field rendering.
\newblock {\em arXiv preprint arXiv:2311.10091}, 2023.

\bibitem{kerbl20233d}
Bernhard Kerbl, Georgios Kopanas, Thomas Leimk{\"u}hler, and George Drettakis.
\newblock 3d gaussian splatting for real-time radiance field rendering.
\newblock {\em ACM Transactions on Graphics}, 42(4):1--14, 2023.

\bibitem{barron2022mip}
Jonathan~T Barron, Ben Mildenhall, Dor Verbin, Pratul~P Srinivasan, and Peter
  Hedman.
\newblock Mip-nerf 360: Unbounded anti-aliased neural radiance fields.
\newblock In {\em Proceedings of the IEEE/CVF Conference on Computer Vision and
  Pattern Recognition}, pages 5470--5479, 2022.

\bibitem{fridovich2022plenoxels}
Sara Fridovich-Keil, Alex Yu, Matthew Tancik, Qinhong Chen, Benjamin Recht, and
  Angjoo Kanazawa.
\newblock Plenoxels: Radiance fields without neural networks.
\newblock In {\em CVPR}, pages 5501--5510, 2022.

\bibitem{wu2024recent}
Tong Wu, Yu-Jie Yuan, Ling-Xiao Zhang, Jie Yang, Yan-Pei Cao, Ling-Qi Yan, and
  Lin Gao.
\newblock Recent advances in 3d gaussian splatting.
\newblock {\em arXiv preprint arXiv:2403.11134}, 2024.

\bibitem{niedermayr2023compressed}
Simon Niedermayr, Josef Stumpfegger, and R{\"u}diger Westermann.
\newblock Compressed 3d gaussian splatting for accelerated novel view
  synthesis.
\newblock {\em arXiv preprint arXiv:2401.02436}, 2023.

\bibitem{fan2023lightgaussian}
Zhiwen Fan, Kevin Wang, Kairun Wen, Zehao Zhu, Dejia Xu, and Zhangyang Wang.
\newblock Lightgaussian: Unbounded 3d gaussian compression with 15x reduction
  and 200+ fps.
\newblock {\em arXiv preprint arXiv:2311.17245}, 2023.

\bibitem{malarzgaussian}
Dawid Malarz, Weronika Smolak, Jacek Tabor, S{\l}awomir Tadeja, and
  Przemys{\l}aw Spurek.
\newblock Gaussian splatting with nerf-based color and opacity.
\newblock {\em arXiv preprint arXiv:2312.13729}, 2023.

\bibitem{yang2024spec}
Ziyi Yang, Xinyu Gao, Yangtian Sun, Yihua Huang, Xiaoyang Lyu, Wen Zhou,
  Shaohui Jiao, Xiaojuan Qi, and Xiaogang Jin.
\newblock Spec-gaussian: Anisotropic view-dependent appearance for 3d gaussian
  splatting.
\newblock {\em arXiv preprint arXiv:2402.15870}, 2024.

\bibitem{saito2023relightable}
Shunsuke Saito, Gabriel Schwartz, Tomas Simon, Junxuan Li, and Giljoo Nam.
\newblock Relightable gaussian codec avatars.
\newblock {\em arXiv preprint arXiv:2312.03704}, 2023.

\bibitem{gao2023relightable}
Jian Gao, Chun Gu, Youtian Lin, Hao Zhu, Xun Cao, Li~Zhang, and Yao Yao.
\newblock Relightable 3d gaussian: Real-time point cloud relighting with brdf
  decomposition and ray tracing.
\newblock {\em arXiv preprint arXiv:2311.16043}, 2023.

\bibitem{chen2024survey}
Guikun Chen and Wenguan Wang.
\newblock A survey on 3d gaussian splatting.
\newblock {\em arXiv preprint arXiv:2401.03890}, 2024.

\bibitem{jankovic2005quadratic}
Vladimir Jankovi{\'c}.
\newblock Quadratic functions in several variables.
\newblock {\em The Teaching of Mathematics}, (15):53--60, 2005.

\bibitem{knapitsch2017tanks}
Arno Knapitsch, Jaesik Park, Qian-Yi Zhou, and Vladlen Koltun.
\newblock Tanks and temples: Benchmarking large-scale scene reconstruction.
\newblock {\em ACM Transactions on Graphics (ToG)}, 36(4):1--13, 2017.

\bibitem{hedman2018deep}
Peter Hedman, Julien Philip, True Price, Jan-Michael Frahm, George Drettakis,
  and Gabriel Brostow.
\newblock Deep blending for free-viewpoint image-based rendering.
\newblock {\em ACM Transactions on Graphics (ToG)}, 37(6):1--15, 2018.

\end{thebibliography}

\end{document}